# Sobre a Evolução do Conceito de Calor e Energia Térmica

Lucas Matheus Alves Gontijo, Clóves Gonçalves Rodrigues

Pontifícia Universidade Católica de Goiás

**Resumo**
*O conceito de energia térmica só foi completamente formulado no fim do século XIX e início do século XX. A procura sobre o entendimento dos fenômenos caloríficos foi o que proporcionou a construção e o desenvolvimento do conceito de energia térmica. Neste trabalho apresentamos a evolução do conceito de calor e energia térmica ao longo do tempo, seguindo uma ordem cronológica e de evolução das ideias e dos princípios em questão. Para isto, foi realizada uma vasta pesquisa bibliográfica sobre o assunto. O tema foi abordado desde os antigos filósofos gregos até o advento da física moderna com a elucidação do fenômeno da radiação térmica.*
***Palavras-chave****: Calor, energia térmica, radiação térmica, flogístico, calórico, história da termodinâmica.*

# About the Evolution of the Concept of Heat and Thermal Energy

**Abstract**
*The concept of thermal energy was only fully formulated in the late 19th and early 20th centuries. The search for the understanding of calorific phenomena was what provided the construction and development of the concept of thermal energy. In this work we present the evolution of the concept of heat and thermal energy over time, following a chronological order and the evolution of the ideas and principles in question. For this, a vast bibliographical research on the subject was carried out. The theme has been addressed from the ancient Greek philosophers to the advent of modern physics with the elucidation of the phenomenon of thermal radiation.*

# INTRODUÇÃO

O conceito de energia térmica só foi completamente formulado no fim do século XIX e início do século XX. A busca sobre o entendimento dos fenômenos caloríficos foi o que proporcionou a construção e o desenvolvimento do conceito de energia térmica. Nota-se historicamente que na medida em que os fenômenos caloríficos foram sendo melhor entendidos ao longo do tempo, o conceito de calor e energia térmica foi paralelamente sendo modificado.

A utilização de fatos históricos, as observações feitas e repassadas geração após geração tem um importante papel no ensino de ciências. No entanto, não se percebe aplicações desta em salas de aula. Segundo Martins: "A necessidade de incorporação de elementos históricos e filosóficos no ensino médio chega a ser praticamente consensual, o que passou a orientar currículos de parcela significativa das licenciaturas. No entanto, os professores do nível médio dificilmente usam esse tipo de conhecimento em suas práticas"[1].

Em geral o estudo de qualquer disciplina (seja matemática, biologia, física, etc.) vem carregado de teorias e conceitos adjacentes à matéria em questão. A introdução de fatos históricos faz o aluno criar em sua mente um tipo de estímulo ao conteúdo, de maneira a facilitar o aprendizado. Assim, alguma introdução histórica, como recurso didático, é sempre bem vinda antes de se iniciar ou mesmo durante o estudo de qualquer conceito teórico em ciências[2]. Mas este não deve ser o único papel da história no ensino de ciências como um todo, e em específico no ensino de física.

A história carrega todo o passado, ou melhor, a construção dos problemas e toda sua complexidade até se chegar aos dias atuais. Infelizmente estas construções históricas têm sido deixadas de lado nos livros didáticos atuais.

O papel da história no ensino de ciências vai além da questão dos fatos. A história é como uma máquina do tempo, que consegue resgatar dificuldades, sofrimentos e alegrias de momentos distantes. Segundo Boff e Barros: "Infere-se que a dificuldade do aprendizado possa ser solucionada mostrando aos estudantes, por meio de discussões abertas, conhecimentos construídos pelo homem ao longo de sua história, como as questões foram respondidas ao longo dos tempos e os caminhos percorridos para sua solução, criando uma situação mais propícia para que os alunos possam contextualizar os conceitos estudados, bem como suas angústias, preocupações, dificuldades e certezas"[3].

Essas discussões podem justificar a importância de introdução de fatos históricos na formação dos professores, mesmo sabendo-se que esses fatos históricos, por si só, não garantem sua utilização de maneira adequada em salas de aula. Segundo Martins: "Assim, a História e Filosofia da Ciência surge como uma *necessidade formativa do professor*, na medida em que pode contribuir para: evitar visões distorcidas sobre o fazer científico; permitir uma compreensão mais refinada dos aspectos envolvendo o processo de ensino aprendizagem da ciência; proporcionar uma intervenção mais qualificada em sala de aula. [...] No entanto, a simples consideração de elementos históricos e filosóficos na formação inicial de professores das áreas das científicas – ainda que feita com qualidade – não garante a inserção desses conhecimentos nas salas de aula do ensino básico, tampouco uma reflexão mais aprofundada, por parte dos professores, do papel da História e Filosofia da Ciência para o campo da didática das ciências. As principais dificuldades surgem quando pensamos na *utilização* da História e Filosofia da Ciência para fins didáticos, ou seja, quando passamos dos cursos de formação inicial para o contexto aplicado do ensino e aprendizagem das ciências"[1].

Neste artigo apresenta-se a evolução do conceito de calor e energia térmica ao longo do tempo, seguindo uma ordem cronológica e de evolução das ideias, dos fenômenos e dos princípios em questão. Para isto, foi realizada uma vasta pesquisa bibliográfica sobre o assunto. O tema foi abordado desde os antigos filósofos gregos, os pré-socráticos, até o advento da mecânica quântica no início do século XX, que teve o seu surgimento ligado à explicação do fenômeno da radiação térmica de corpo negro.

É muito bem sabido que livros textos sobre evolução dos conceitos da física e sobre história da física também apresentam a ordem cronológica e a evolução de ideias e princípios científicos. No entanto, tais livros abordam diferentes conceitos físicos e não apenas um único tema, não entrando em muitos detalhes sobre um determinado assunto. No presente artigo discorremos sobre um único tema (a evolução do conceito de calor e energia térmica) abordando mais detalhadamente o assunto e com o cuidado de explorar os contextos nessa construção para além de uma mera cronologia de fatos e personagens, apresentando de forma clara e didática o processo gradativo e progressivo de transformação do conceito de calor.

## IDEIAS DE FILÓSOFOS ANTIGOS SOBRE A ORIGEM E A COMPOSIÇÃO DA MATÉRIA

Segundo o poeta grego da antiguidade Hesíodo, que viveu no século VIII a.C., a tarefa de criar os homens e todos os animais foi dada ao titã Prometeu (veja ilustração na Fig. 1) e a seu irmão Epimeteu[4]. Epimeteu ficou encarregado da obra e Prometeu de supervisioná-la. Epimeteu atribuiu a cada animal dons variados como força, coragem, sagacidade, rapidez; asas a um, garras a outro, carapaça protetora a um terceiro, e assim por diante. No entanto, quando chegou a vez do homem criou-o do limo da terra. Mas como Epimeteu gastou todos os recursos com os outros animais, recorreu a seu irmão Prometeu. Prometeu, então, roubou o fogo dos deuses e o deu aos homens. Isto assegurou a superioridade dos homens sobre os outros animais. Todavia o fogo era exclusivo dos deuses. Prometeu foi, então, castigado pelos Deuses.[5]

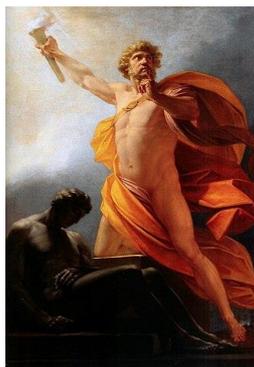

**Figura 1:** Prometeu leva o fogo à humanidade (por Heinrich Friedrich Füger, 1817).[6]

Assim, pode-se dizer que inicialmente, desde a antiguidade, o fogo e os fenômenos caloríficos estavam estreitamente relacionados, e desta forma a procura pelo entendimento do que era o fogo e o seu "poder" (propriedades) foi uma das tarefas iniciais no campo de entendimento dos fenômenos caloríficos. Isto levou também a uma relação íntima entre o fogo e a busca pelo elemento essencial da Natureza.

Os filósofos pré-socráticos tinham opiniões divergentes quanto a este elemento essencial na Natureza[7]. Este elemento era chamado de arché. Para os filósofos pré-socráticos, a *arché* ou *arqué*, seria o elemento que deveria estar presente em todos os momentos da existência de todas as coisas do mundo. Nesta linha destacamos os seguintes filósofos:

- **Tales de Mileto: a água**
  Para Tales de Mileto, Fig. 2, a arché seria a água. Provavelmente ao visitar o Egito, Tales observou que os campos de plantações após serem inundados pelo rio Nilo ficavam fecundos. Observou também que o calor necessita de água, que o morto resseca, que a natureza é úmida, que os germens são úmidos, que os alimentos contêm seiva, e assim concluiu que o princípio de tudo era a água. Tales considerava a água como arché em todos os seus possíveis estados físicos. Tudo, então, seria a alteração dos diferentes graus da água.

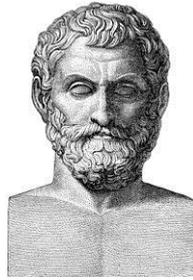

**Figura 2:** Busto de Tales de Mileto (c.624 – 546 a.C.).[8]

- **Anaximandro de Mileto: o Ápeiron**

  Anaximandro tinha um argumento contra Tales: o ar é frio, a água é úmida, e o fogo é quente, e essas coisas são antagônicas entre si. Portanto, o elemento primordial não poderia ser um dos elementos visíveis, teria que ser um elemento neutro, que está presente em tudo, mas é invisível. Considerava o *infinito* como o princípio das coisas, e o chamou de Éter. Anaximandro acreditava que as coisas nascem do infinito através de um processo de separação dos contrários (seco-úmido). Anaximandro afirmou que os primeiros animais nasceram no elemento líquido e, pouco a pouco, vieram para o ambiente seco, mudando o seu modo de viver por um processo de adaptação ao ambiente. Notamos aqui uma incrível semelhança com as teorias evolucionistas.

- **Anaxímenes de Mileto: o ar**

  Anaxímenes de Mileto um discípulo de Anaximandro, Fig. 3, discordava que os contrários poderiam gerar várias coisas. Colocou o ar como Arché, pela necessidade vital deste para os seres vivos e porque segundo ele o ar, melhor que qualquer outra coisa, sofre variações. A rarefação e condensação do ar formam o mundo. A alma é ar, o fogo é ar rarefeito; quando acontece uma condensação, o ar se transforma em água, se condensa ainda mais e se transforma em terra, e por fim em pedra.

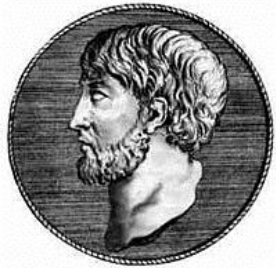

**Figura 3:** Rosto de Anaxímenes de Mileto (c. 570 – c. 500).[9]

- **Xenófanes de Cólofon: a terra**

  O elemento primordial para Xenófanes, Fig. 4, era a terra. Através do elemento terra desenvolve sua cosmologia. Sua filosofia tinha sua lógica, pois, afinal, tudo o que existe começa na terra e tudo volta para a terra, tanto animais quanto plantas. Aos que questionavam o porquê da terra se a maior parte do planeta era feita de água, a resposta era que o fundo do oceano era feito de terra.

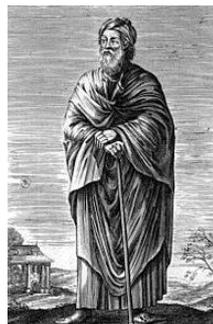

**Figura 4:** Ilustração de Xenófanes de Cólofon (ca. 570 a.C. – 475 a.C.).[10]

- **Heráclito de Éfeso: o fogo**
  Heráclito atribuiu ao fogo como princípio de todas as coisas. Segundo Heráclito o fogo transforma-se em água, sendo que uma metade retorna ao céu como vapor e a outra metade transforma-se em terra. Sucessivamente, a terra transforma-se em água e a água, em fogo. Mas Heráclito era mobilista e afirmava que todas as coisas estão em movimento como um fluxo perpétuo. Ou seja, usa o fogo apenas como símbolo de todo este movimento.

- **Pitágoras de Samos: o número**
  Os pitagóricos interessavam-se pelo estudo das propriedades dos números. Para os pitagóricos o número (sinônimo de harmonia) era considerado como essência das coisas. O número seria constituído da soma de pares e ímpares, expressando as relações que se encontram em permanente processo de mutação. Teriam chegado à concepção de que todas as coisas são números.

- **Empédocles de Agrigento: os quatro elementos**
  Empédocles, Fig. 5, acreditava que a natureza possuía quatro elementos básicos: a terra, o ar, o fogo e a água. Não é certo, portanto, afirmar que "tudo" muda. Basicamente, nada se altera. O que acontece é que esses quatro elementos diferentes simplesmente se combinam e depois voltam a se separar para então se combinarem novamente. O que unia e desunia os quatro elementos eram dois princípios: o amor e o ódio. Os quatro elementos e os dois princípios seriam eternos e imutáveis, mas as substâncias formadas por eles seriam pouco duradouras. O amor agiria como força de atração e união, o ódio como força de dissolução. Em quatro fases, existe a alternância dos dois. Estabelece um ciclo, com a tensão da convivência dessas forças motrizes.

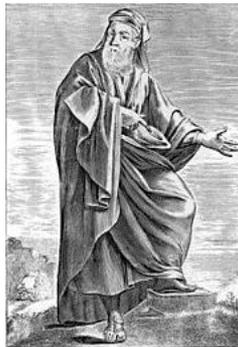

**Figura 5:** Ilustração de Empédocles de Akragas (atual agrigento) (c. 490–c. 430).[11]

- **Anaxágoras de Clazomena: as homeomerias**
  Anaxágoras achava que a natureza era composta por uma infinidade de partículas minúsculas, invisíveis a olho nu. Assim, em tudo existia um pouco de tudo. Anaxágoras chamou as infinitas partículas de homeomerias, ou sementes invisíveis, que diferiam entre si nas qualidades. Todas as coisas resultariam da combinação das diferentes homeomerias.

- **Demócrito de Abdera e Leucipo de Mileto: os átomos**
  Os atomistas, Fig. 6, seguiram a linha de que a natureza era composta por partículas infinitas. Diziam que tudo que realmente existia era constituído de átomos e de vazio, sendo o vazio o espaço entre os átomos. Consideravam que nada pode surgir do nada, assim, os átomos eram eternos, imutáveis e indivisíveis. Os átomos eram irregulares e podiam ser combinados para dar origem aos corpos mais diversos. Para os atomistas o fogo, por exemplo, seria constituído por átomos esféricos.

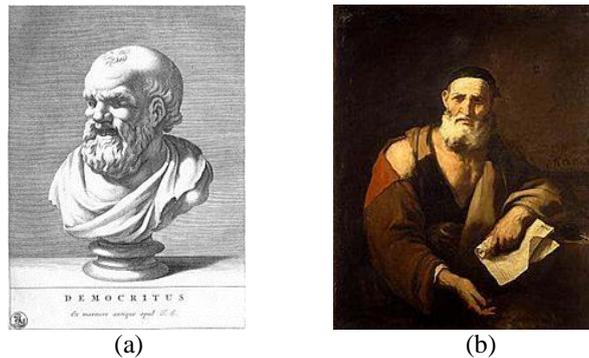
**Figura 6:** (a) Demócrito de Abdera (470 a.C. – 370 a.C.), (b) Leucipo de Mileto (c. 460 – c. 370).[12]

● **Platão**

O filósofo grego Platão, Fig. 7, defendia que a Natureza era composta por quatro elementos essenciais os quais estavam relacionados com os poliedros regulares pitagóricos: fogo→tetraedro, ar→octaedro, terra→hexaedro, e água→icosaedro. Havia ainda, para Platão, um quinto elemento pitagórico, o dodecaedro, o qual representava o Universo.

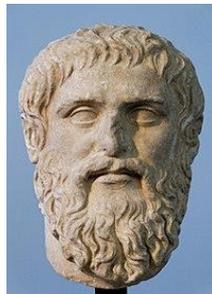
**Figura 7:** Platão (Atenas, 428/427 – Atenas, 348/347 a.C).[13]

● **Aristóteles**

Aristóteles, Fig. 8, também era defensor de que a Natureza era composta por quatro elementos essenciais. Para Aristóteles os quatro elementos essenciais eram: o quente, o seco, o frio, e o úmido. Os elementos empedoclianos eram decorrentes de uma combinação desses elementos. Dessa forma: ar: quente↔úmido; fogo: quente↔seco; terra: frio↔seco; água: frio↔úmido.

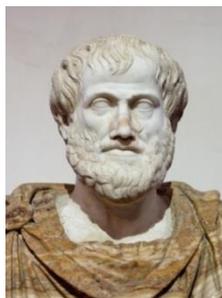
**Figura 8:** Aristóteles (Estagira, 384 a.C. – Atenas, 322).[14]

● **Filósofos chineses**

Filósofos chineses também propuseram elementos básicos na busca de uma compreensão para o Universo. O filósofo chinês Zou Yan, Fig. 9, por exemplo, considerava que os elementos essenciais da Natureza eram: fogo, terra, metal, água, e madeira. Estes cinco elementos eram governados pelo dualismo básico dos princípios cósmicos do "yang" (penetrante, macho, ativo, céu) e do "yin" (absorvente, fêmea, passivo, terra,). Assim, esses cinco elementos não eram apenas meras substâncias.[15]

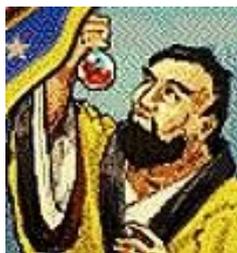

**Figura 9:** Ilustração de Zou Yan (c. 360 – c. 260).[16]

- **Filósofos hindus**

Os filósofos hindus também acreditavam na existência de elementos essenciais na composição do Universo. Estes elementos eram considerados manifestações da alma, chamada de "atman". Esses filósofos hindus listavam cinco *elementos*-manifestações que estavam conectados aos sentidos humanos da seguinte forma: *terra*↔olfato, *éter*↔audição, *fogo*↔visão, *ar*↔tato, *água*↔paladar.[17]

- **Filósofos árabes**

Os árabes também tiveram um papel importante, principalmente a partir do século VIII. Nesta empreitada, destaca-se que eruditos árabes traduziram textos hindus e gregos. O alquimista árabe Abu Muça Jabir Ibne Haiane, mais conhecido como Jabir, Fig. 10(a), considerava que os quatro elementos empedoclianos eram associados de tal forma que o resultado era somente dois elementos: *i*) o "enxofre" responsável pela combustão, ou seja, o fogo, e *ii*) o "mercúrio" que era o responsável pelas propriedades metálicas e era associado ao elemento "água", ou seja, os líquidos. Jabir verificou, em uma de suas inúmeras experiências, que a cal formada no processo de calcinação de um metal era mais pesada que o metal antes de ser queimado[18]. O alquimista e médico suíço Theophrastus Philippus Aureolus Bombastus von Hohenheim, mais conhecido como Paracelso, Fig. 10(b), incorporou o sal como o responsável pela solidificação (terra) das substâncias. Sua teoria, no entanto, não explicava porque na combustão de metais o peso do corpo não era conservado.

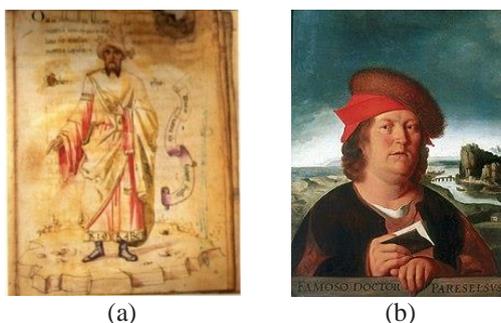

(a)  (b)
**Figura 10:** (a) Abu Musa Jabir Ibn Hayyan (c. 721-c. 815), (b) Paracelso (1493-1541).[19]

## O FLOGÍSTICO

Em continuidade com a formulação de ideias sobre a matéria, o químico alemão Johann Joachim Becher, Fig. 11, no ano de 1669, apresentou em seu livro "Física Subterrânea" a ideia de que os corpos sólidos eram classificados em três espécies "terra": a "terra mercurial", a "terra gordurosa" (combustível), e a "terra vitrificável", a qual era análoga ao "sal" de Paracelso. Segundo Becher, a "terra gordurosa" era a responsável pelo processo da combustão porque era esta espécie de "terra" que era liberada quando um corpo sofria combustão (queima).

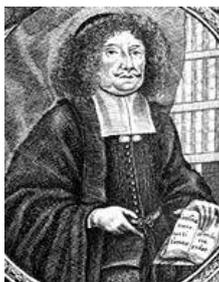

**Figura 11:** Johann Joachim Becher (1635-1682): autodidata, adquiriu vastos conhecimentos, não só químicos, mas em muitas outras áreas, chegando a dominar dez idiomas.[20]

O químico e médico alemão Georg Ernst Stahl, Fig. 12, apresentou uma teoria mais abrangente sobre a "terra gordurosa" (também chamada de terra "combustível") no ano de 1697 em seu livro "*Experimenta, observationes, animadvertiones chymical et physical*". Neste livro Stahl utiliza o termo grego "phlogistós" (flogístico) que significa consumido pelo fogo. Quando algum tipo de metal era retirado de uma cal metálica por um processo de queima de carvão, dizia-se que a cal absorvia flogístico do carvão, e quando o metal era oxidado (calcinado) o metal perdia flogístico. E para explicar porque os metais tinham o seu peso aumentado quando se oxidavam admitia-se que o flogístico tinha peso negativo por ser considerado "leve".[21]

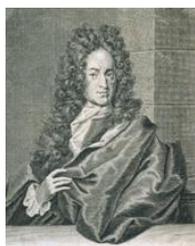

**Figura 12:** Georg Ernst Stahl (1660-1734): nasceu em Ansbach, na Baviera.[22]

Assim, a ideia do flogístico de Stahl tomou conta do pensamento de praticamente todos os cientistas do século XVIII. Nesta época, os gases que eram descobertos tinham o seu nome associado ao flogístico. No ano de 1772, por exemplo, o químico e médico escocês Daniel Rutherford, Fig. 13(a), observou em experimentos em seu laboratório que no ar totalmente viciado pela sua queima ou pela respiração de camundongos existia, além do ar fixo (hoje chamado de gás carbônico), um outro tipo de ar que não podia ser respirado. Este ar foi então chamado de "ar flogisticado", e que não servia para a combustão e respiração, processos que para serem realizados necessitavam do flogístico. Este ar flogisticado era na verdade o que hoje chamamos de nitrogênio. Daniel Rutherford descreveu também o oxigênio, que denominou de "ar vital".[23]

Em 1774 o químico inglês Joseph Priestley, Fig. 13(b), notou que uma cal de cor avermelhada desprendia um gás quando aquecida por uma lente. A cal era óxido de mercúrio ($HgO$) que na época recebeu o nome de "mercurius calcinatus per se". Priestley também notou que, colocando este gás desprendido pela cal em um recipiente e introduzindo uma vela acesa nele, a chama da vela ficava num tom de cor bem vivo. Como este gás tinha propriedades justamente opostas às do "ar flogisticado", Priestley denominou este gás de "ar deflogisticado" ou "ar perfeito"[24]. Sabe-se hoje que este gás nada mais é que o oxigênio. Assim, mais um tipo de gás tinha o seu nome associado ao flogístico. Ressalta-se aqui que em 1722 o químico e farmacêutico sueco Karl Wilhelm Scheele, Fig. 13(d), também havia descoberto este mesmo tipo de gás. Scheele descobriu este gás aquecendo substâncias oxigenadas e deu o nome a esse gás de "ar de fogo" ou "ar empíreo" pelo fato desse gás favorecer à combustão[25]. Normalmente os livros textos dão a Joseph Priestley o título de descobridor do oxigênio, uma vez que a sua obra foi a primeira a ser publicada no ano de 1774.

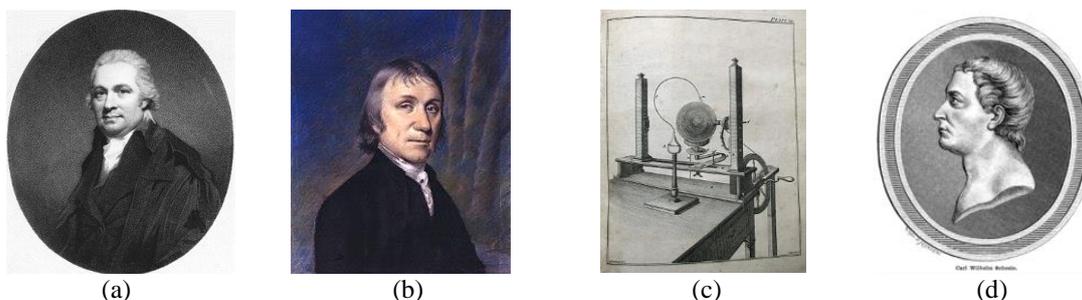

(a) (b) (c) (d)

**Figura 13:** (a) Daniel Rutherford (1749-1819): conhecido por ter descoberto e isolado o elemento químico nitrogênio descrevendo, em 1772, algumas de suas propriedades. (b) Joseph Priestley (1733-1804) foi um teólogo, clérigo dissidente, filósofo natural, educador, teórico e político britânico que publicou mais de 150 obras; (c) Máquina elétrica de Priestley para amadores experimentalistas, ilustrada na primeira edição de "Familiar Introduction to Electricity" em 1768, que ele comercializou sem sucesso com seu irmão Timothy; (d) Karl Wilhelm Scheele (1742-1786): Scheele nasceu em Stralsund, na Pomerânia hoje Alemanha, na época uma província sueca.[26]

Em 1781, Priestley ao explodir em seu laboratório uma centelha elétrica dentro de uma garrafa que continha em seu interior uma mistura de ar inflamável (hidrogênio) e ar deflogisticado (oxigênio) notou que minúsculas gotículas de água (uma espécie de orvalho) havia se formado na superfície interna da garrafa. Priestley não deu muita importância ao fenômeno concluindo somente que "o ar comum deposita sua mistura quando flogisticado". O químico e físico inglês Henry Cavendish, ainda no ano de 1781, repetiu a experiência realizada por Priestley, porém, Cavendish notou algo que Priestley não havia percebido: as minúsculas gotículas de água que se formaram na face interna da garrafa era água pura. Como Cavendish era adepto do chamado flogístico stahliano conclui que os "ares" dentro da garrafa eram água sem flogístico e água com flogístico[27]. Assim, sua crença no flogístico impediu Cavendish de descobrir o que seria na verdade a composição da água.

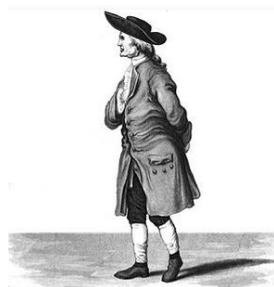

**Figura 14:** Henry Cavendish (1731-1810). Henry Cavendish nasceu em outubro de 1731 em Nice, França, então parte do Reino da Sardenha onde sua família morava na época. Morreu em fevereiro de 1810 em Londres.[28]

# O CALÓRICO

Em 1772, realizando experiências sobre combustão, o químico francês Antoine Laurent Lavoisier, Fig. 15(a), começou a questionar a ideia do flogístico. Com estas experiências Lavoisier chegou à conclusão que quando queimadas as substâncias ganham peso pela absorção de "ar deflogisticado". Lavoisier denominou este ar de oxigênio (Parks, 1939). O nome oxigênio deriva de dois vocábulos gregos: "oxys" (ácido, literalmente picante, em alusão ao sabor dos ácidos) e "gonos" (produtor, literalmente gerar), porque na época em que foi utilizada esta denominação acreditava-se, equivocadamente, que todos os ácidos necessitavam de oxigênio para a sua composição. Lavoisier concluiu que a causa do calor era um "fluido imponderável" e chamou este fluido de "matéria de fogo", o qual dependendo da quantidade poderia formar um dos três estados da matéria: o sólido, o líquido e o gasoso[29]. É notável ressaltar que o nome "matéria de fogo" foi substituído por "calórico" no livro que tinha como título: "Método da Nomenclatura Química". Este livro foi publicado no ano de 1787 por Lavoisier e pelos químicos franceses: Barão Louis Bernard Guyton de Morveau, Conde Antoine François Fourcroy e Conde Claude Louis Berthollet, Fig. 15.

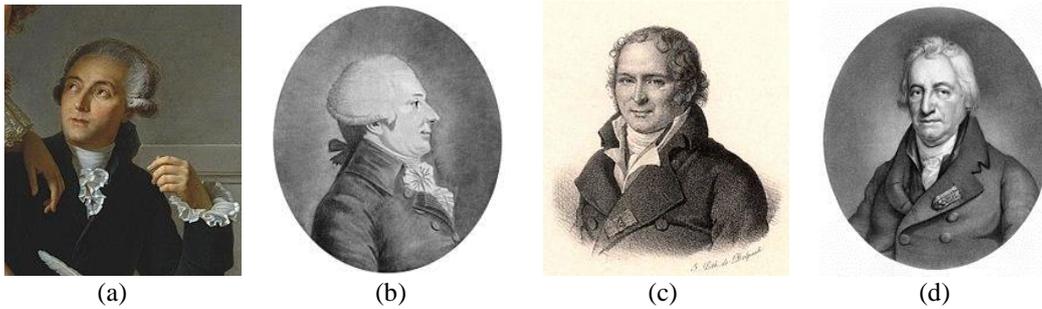

**Figura 15:** (a) Antoine Laurent Lavoisier (1743-1794): nascido em família rica de Paris, herdou grande fortuna com idade de cinco anos quando sua mãe faleceu. (b) Barão Louis Bernard Guyton de Morveau (1737-1816): considerado o primeiro a produzir um método sistemático de nomenclatura química. (c) Conde Antoine François Fourcroy (1755-1809): descobriu o fosfato de magnésio, desenvolveu estudos sobre gelatina e albumina, melhorou o processo de separação do estanho e cobre. (d) Conde Claude Louis Berthollet (1748-1822): indicado em 1780 para membro da Academia das Ciências da França, eleito membro da Royal Society em 1789 devido à publicação de numerosos trabalhos científicos de alta qualidade.[30]

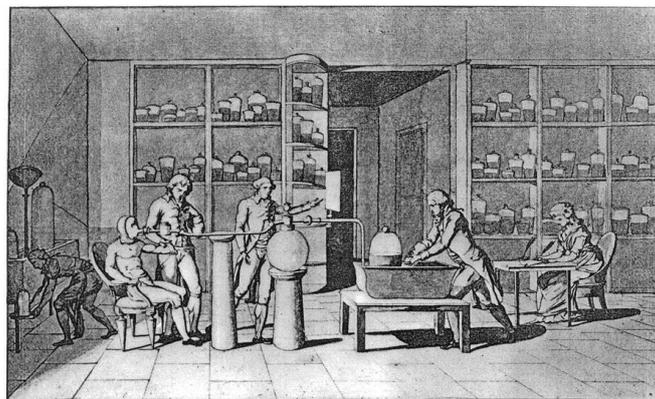

**Figura 16:** Madame Lavoisier (sentada à direita) ajudando o marido em sua pesquisa científica da respiração humana.[31]

Em 1789 Lavoisier publica seu livro intitulado "Tratado Elementar de Química". Neste livro Lavoisier define o calórico como sendo uma espécie de fluido imponderável que tem a capacidade de penetrar e fluir através de qualquer substância. Ainda neste texto Lavoisier defende que a água era composta de oxigênio e hidrogênio com base em seus experimentos datados de 1783, concluindo que a água não era um dos elementos gregos primordiais da Natureza. Destaca-se que Lavoisier era bem cuidadoso quanto à descrição dos equipamentos que utilizava em seus experimentos[32], como o calorímetro de gelo ilustrado na Fig. 17. Outro importante resultado apresentado por Lavoisier neste livro foi o famoso enunciado da sua lei (ou princípio) de conservação da massa: "Na Natureza nada se perde e nada se cria, há somente alterações e modificações". Esta era a base da teoria da combustão de Lavoisier.

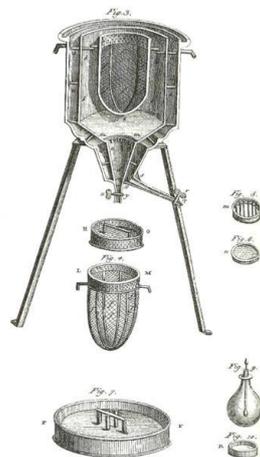

**Figura 17:** Ice-calorimeter de Lavoisier.[33]

Deve-se fazer a ressalva que o princípio de conservação da massa havia sido enunciado 14 anos antes de Lavoisier pelo cientista russo Mikhail Lomonossov, Fig. 18(a). A lei da conservação das massas foi publicada pela primeira vez em um ensaio de Lomonosov no ano de 1760[34]. No entanto, a obra não repercutiu na Europa Ocidental, cabendo a Lavoisier o papel de tornar mundialmente conhecido o que hoje se chama lei de Lavoisier.[35]

Deve ser notado que a estrutura química da água como sendo $H_2O$, só foi melhor estabelecida quando no ano de 1811 o conde italiano Amedeo Avogadro, Fig. 18(b), apresentou a sua hipótese de distinção entre "molécula e "átomo".[36]

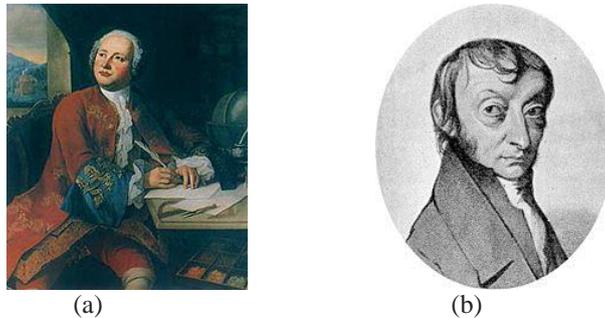

(a)          (b)

**Figura 18:** (a) Mikhail Vasilyevich Lomonosov (1711-1765): fez diversas descobertas foram depois confirmadas por cientistas da Europa ocidental. Um dos principais exemplos é a experiência de calcinar metais em recipientes fechados, realizada em 1760, que seria repetida por Antoine Lavoisier, 13 anos depois, servindo de base para a descoberta da Lei da Conservação da massa. (b) Amedeo Avogadro (1776-1856): Trabalhou na Universidade de Turim como responsável pela cadeira de física por 30 anos, período que publicou boa parte de suas obras científicas.[37]

Infelizmente, Lavoisier foi guilhotinado em 8 de maio de 1794 por ter sido um rendeiro geral. Os rendeiros gerais eram muito impopulares por causa das cobranças de impostos que muitos praticavam. Todos foram condenados à guilhotina. Num total de vinte e oito rendeiros gerais Lavoisier foi o quarto a ser executado. No dia posterior desse julgamento, Joseph-Louis de Lagrange (1736-1813), um importante matemático, contemporâneo de Lavoisier disse: "Não bastará um século para produzir uma cabeça igual à que se fez cair num segundo".[38]

Os cientistas defensores do flogístico usavam a ideia de ganho ou de perda de uma espécie de "partícula ígnea". Esta foi a mesma base ideológica adotada pelos cientistas defensores do calórico. Os cientistas do final do século XVIII e das décadas iniciais do século XIX acreditavam que a retirada ou adição do fluido calórico a um corpo, tal como havia definido Lavoisier, era o motivo do porque os corpos aumentavam ou diminuíam as suas dimensões quando a sua temperatura era aumentada ou diminuída, respectivamente. Porém, no ano de 1776, o geólogo, filósofo e meteorologista suíço Jean-André Deluc, Fig. 19(a), mostrou em um de seus experimentos, que a água aumentava o seu volume com a diminuição da temperatura para valores menores que aproximadamente 4ºC, Fig. 19(b), contradizendo a teoria do fluido calórico, já que o esperado por esta teoria era que o volume diminuísse ao invés de aumentar.[39]

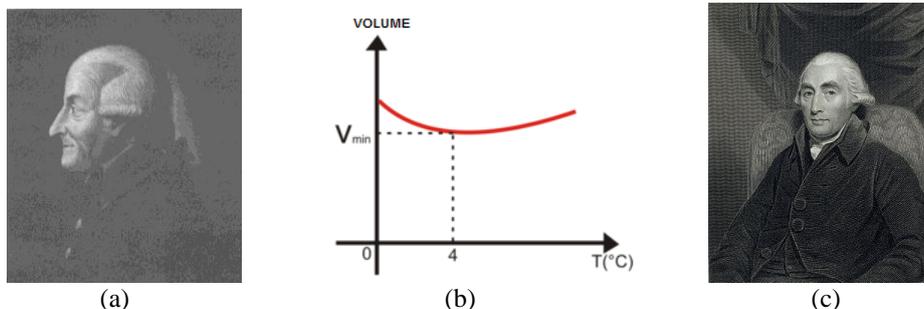

(a)          (b)          (c)

**Figura 19:** (a) Jean-André Deluc (1727-1817): inventou e aperfeiçoou alguns instrumentos de medida como, por exemplo, o barômetro e o higrômetro. (b) Dilatação anômala da água.[40]

Devemos destacar aqui a importante contribuição feita por Joseph Black, Fig. 19(c), um pouco antes dos trabalhos de Jean Deluc. Antes de 1776, mais especificamente no ano de 1761, Black observou que fornecendo calor ao gelo quando este está em seu ponto de fusão não há um aumento de temperatura da mistura água/gelo, no entanto, a quantidade de água na mistura aumentava. Em um segundo experimento Black verificou também que fornecendo calor a água em seu ponto de ebulição não há um aumento na temperatura da mistura água/vapor, entretanto, a quantidade de vapor aumentava. Black concluiu dessas suas observações que o calor fornecido se combinou com as partículas de gelo no primeiro experimento e com as de água em ebulição no segundo experimento, tornando-se calor latente[41]. Estas experimentações de Joseph Black enfraqueciam a ideia de calórico, uma vez que nestes casos mesmo fornecendo continuamente calor ao corpo a temperatura deste não aumentava.

Outro ponto negativo quanto à teoria do calórico era que este era um fluido que nenhum cientista tinha conseguido medir o seu peso em todos os experimentos realizados até aquele momento.

## O CALOR COMO FORMA DE MOVIMENTO

Enfatizamos aqui a existência de alguns pensadores e cientistas, que tiveram ideias alternativas ao flogístico e ao calórico muito antes dos cientistas já citados para explicar os fenômenos caloríficos. Por exemplo, no ano de 1620, o filósofo inglês Francis Bacon, Fig. 20(a), declarou que o calor era uma forma de movimento[42]. Esta afirmação de Bacon, no entanto, não se firmava em um experimento analisado por ele, podendo ser considera mais uma declaração filosófica que científica. A percepção do calor como forma de movimento também foi defendida, em 1665, pelos ingleses Robert Boyle, Fig. 20(b) e Robert Hooke, Fig. 20(c), e em 1722 pelo filósofo John Locke, Fig. 20(d). Voltamos a citar novamente o cientista russo Mikhail Lomonosov, que no ano de 1740, enunciou: "A causa do calor e do frio tem como origem o movimento mútuo de imperceptíveis e diminutas partículas".

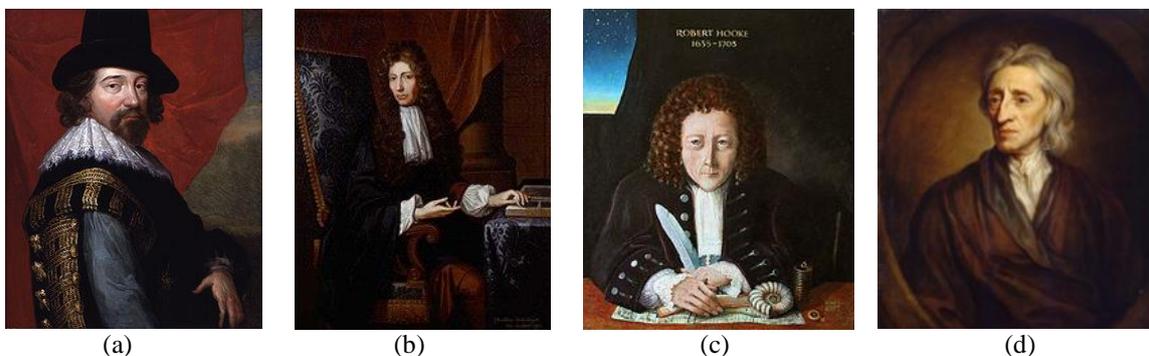

(a) (b) (c) (d)

**Figura 20:** (a) Francis Bacon (1561-1626): foi um político, filósofo, cientista e ensaísta inglês. Considerado por muitos o fundador da ciência moderna. (b) Robert Boyle (1627-1691): uma de suas mais importantes descobertas foi a Lei de Boyle-Mariotte, onde ele dizia que o volume de um gás varia de forma inversamente proporcional com a pressão. (c) Robert Hooke (1653-1703): dentre diversos trabalhos, formulou a teoria do movimento planetário, a primeira teoria sobre as propriedades elásticas da matéria, descreveu a estrutura celular da cortiça. (d) John Locke (1631-1704): ficou conhecido como o "pai do liberalismo".[43]

No ano de 1785 o médico escocês George Fordyce, Fig. 21(a), realizou experiências para determinar se havia mudança de peso na água quando esta mudava de estado líquido para sólido.

E a partir de 1787 o físico e inventor anglo-americano Sir Benjamin Thompson, mais conhecido como Conde de Rumford, Fig. 21(b), realizou inúmeras experiências com objetivos semelhantes. Numa delas Rumford fez o seguinte: em um frasco colocou água e em um segundo frasco idêntico ao primeiro colocou mercúrio, vedando os dois frascos. Em seguida pesou os frascos a uma temperatura de 61ºF e depois a uma temperatura de 30ºF. Como a água e o mercúrio possuem calores específicos bem diferentes era de se esperar que fosse obtida uma diferença na medida de peso em cada uma das temperaturas. Porém, Rumford não encontrou nenhuma diferença na medida do peso nos frascos.[44]

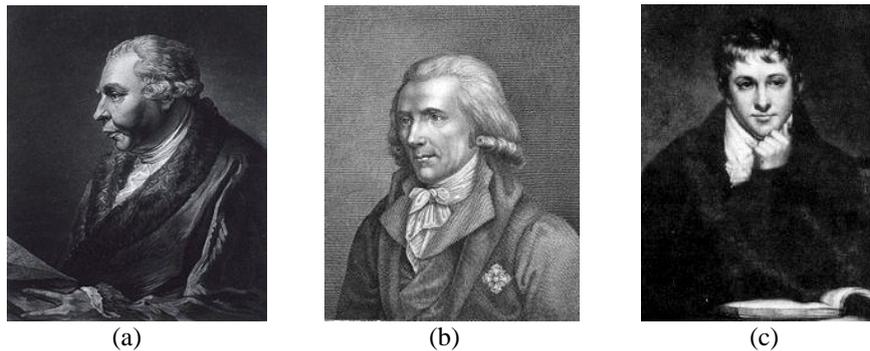

(a)  (b)  (c)

**Figura 21:** (a) Conde Rumford (1753-1814): foi o primeiro cientista a dar evidências que o calor não era uma substância. (b) George Fordyce (1736-1802): professor de medicina e químico, membro da Royal Society e membro do Royal College of Physicians. (c) Sir Humphry Davy (1778-1829): tornou-se famoso por suas experiências sobre a ação fisiológica de alguns gases; teve em seu laboratório como assistente nada menos o genial que Michael Faraday.[45]

Ao inspecionar a produção de canhões feitos de bronze na Baviera, quando era Ministro da Guerra, Rumford notou que blocos maciços de bronze ficavam incandescentes quando a broca perfuradora utilizada para fazer o furo ia desgastando o metal. O cientista britânico notou a ausência de troca de substâncias durante o processo, mesmo havendo aquecimento de ambos os materiais, o que colocava em cheque a ideia que o calor era uma substância. Em uma de suas experiências Rumford introduziu na água um canhão que ia ser perfurado, com o objetivo de que o calor produzido pela broca em atrito com o canhão fosse transferido para a água. Foram utilizados cavalos conectados ao eixo da broca para fazer a broca girar, como ilustrado na Fig. 22. Depois de algumas horas da broca perfurando o canhão, a água começou a ferver. Rumford conclui que fora o movimento dos cavalos que havia fervido a água através do atrito da broca perfurando o canhão. Depois de realizar e analisar vários experimentos neste sentido, Rumford chegou à conclusão que não havia como explicar o fenômeno abordando o calor como um fluido deduzindo, então, que a energia que aquecia os materiais tinha como origem a energia mecânica das brocas em rotação.

Ao correlacionar a equivalência da energia motriz com o calor, Rumford deixou claro que o calor é uma forma de energia, e se tornou um dos primeiros cientistas a analisar a equivalência entre calor e energia mecânica, não podendo esta análise ser creditada unicamente a Joule, do qual falaremos na próxima seção. As pesquisas de Rumford foram publicadas nos anos de 1798 e 1799, onde ele enunciou que: "o calor é uma forma de movimento e não tem peso".

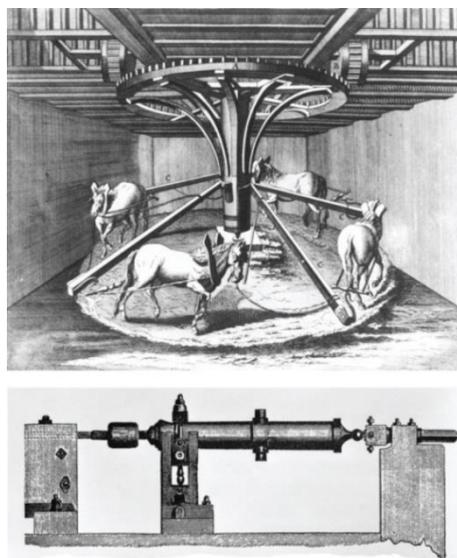

**Figura 22:** Ilustração de um aparato utilizado para perfurar sólidos metálicos.[46]

Em 1799, o químico inglês Sir Humphry Davy, Fig. 21(c), realizou experiências sobre a produção de calor por meio de atrito, chegando à mesma conclusão que Rumford, ou seja, que o calor não tem peso e é uma forma de movimento.[47]

## CALOR E TERMODINÂMICA

Outro grande passo no estudo sobre as trocas de calor foi dado pelo filósofo e físico suíço Pierre Prévost, Fig. 23, que apresentou no ano de 1791 a chamada lei das permutas de Prévost. Segundo esta teoria, um corpo emite e absorve energia radiante em proporções iguais quando está em equilíbrio térmico com a sua vizinhança, e a sua temperatura se mantém constante. No entanto, se o corpo não estiver à mesma temperatura da vizinhança, existirá um fluxo livre de energia entre o corpo e a sua vizinhança devido à uma emissão e absorção desiguais. Suas ideias foram publicadas em 1792 no livro intitulado "Pesquisas Físico-Mecânicas sobre o Calor".[48]

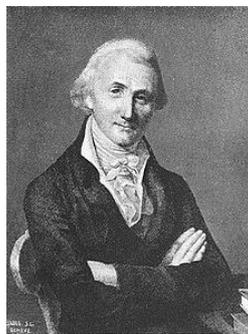

**Figura 23:** Pierre Prévost (1751-1839): argumentou que todos os corpos irradiam calor, por mais quentes ou frios que sejam.[49]

No ano de 1800 o astrônomo alemão Sir William Herschel realizou experiências nas quais observou que as temperaturas das cores do espectro solar aumentavam à medida que se aproximavam da extremidade vermelha do espectro, Fig. 24(a). Neste experimento Herschel descobriu que a temperatura mais alta estava além da cor vermelha do espectro decomposto. Deve-se comentar que a descoberta foi, de certa forma, acidental: Herschel colocou um termômetro de controle justamente na posição do infravermelho, e quando comparou a temperatura dos termômetros que estavam dispostos em cada uma das cores com o termômetro de controle viu que a temperatura deste era a maior. Com base neste fato Herschel apresentou a hipótese de existência de raios além do vermelho os quais chamou de "raios infravermelhos". Em experiências futuras Herschel mostrou que esses raios infravermelhos também podiam ser refratados e refletidos da mesma forma que os raios luminosos. É interessante ressaltar que o físico alemão Johann Wilhelm Ritter, Fig. 24(b), estendeu os experimentos de Herschel para além da cor violeta do espectro solar e descobriu no ano seguinte, em 1801, a existência dos chamados "raios ultravioletas".[50]

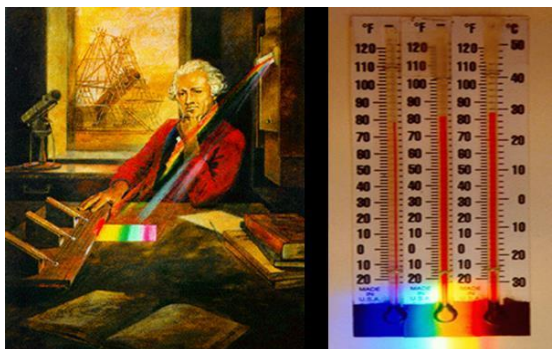 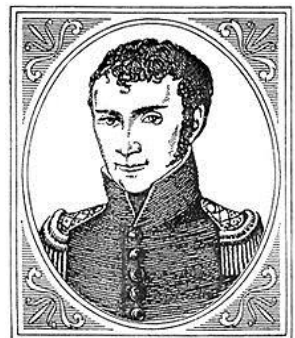

(a)             (b)

**Figura 24:** (a) Experiência de William Herschel (1738-1822). (b) Johann Wilhelm Ritter (1776-1810): fundador da teoria eletroquímica.[51]

Em mais um de seus experimentos, datado de 1804, o Conde Rumford inseriu um termômetro em um balão com ar rarefeito. Após vedar o balão com o termômetro dentro, colocou-o dentro de um recipiente com água quente e observou uma pequena variação para mais na temperatura marcada pelo termômetro. Dessa forma Rumford chegou à conclusão que o calor podia se propagar no vácuo. Em outras experiências, verificou que superfícies que emitiam calor mais fracamente eram as que refletiam mais intensamente. Observou também que a temperatura de um corpo escurecido se relacionava com a intensidade da radiação solar. O matemático e físico escocês Sir John Leslie, Fig. 24(b), também obteve no ano de 1804, resultados similares aos obtidos por Rumford. Em sua experiência sobre calor radiante, Leslie usou um recipiente cúbico contendo água quente. Uma face do cubo era de metal altamente polido, duas faces de metal fosco (cobre) e uma face pintada de preto. Ele mostrou que a radiação era maior na face preta e desprezível na face polida. O aparelho ficou conhecido como "cubo de Leslie". A condução do calor em barras metálicas foi também estudada experimentalmente no mesmo ano, em 1804, pelo físico francês Jean-Baptiste Biot, Fig. 24(c). Biot apresentou como resultados relativos a este experimento a diferença existente entre a radiação externa e a interna em um meio material, entendendo-se por radiação interna à transferência de calor por condução dentro do sólido enquanto que a externa se refere à transferência de calor existente entre a superfície do sólido e o meio.[52]

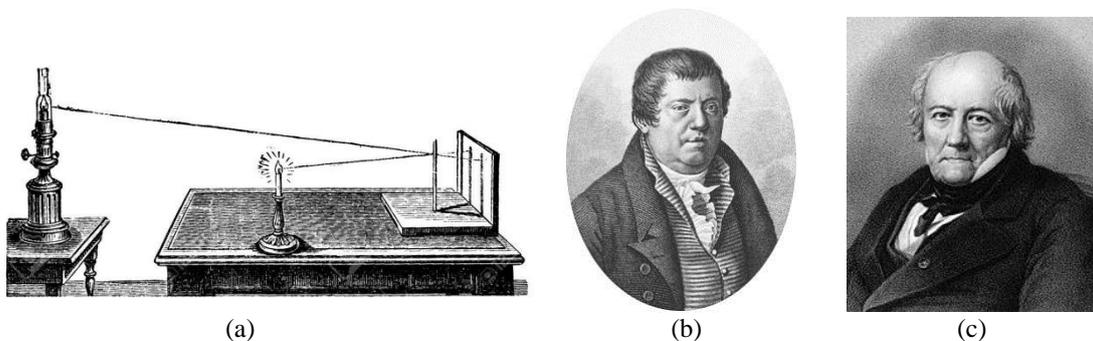

(a) (b) (c)

**Figura 25:** (a) Tipos de sombras do fotômetro Rumford. Fonte: Industrial encyclopedia E. -O. Lami – 1875. (b) Sir John Leslie (1766-1832): fez o primeiro relato moderno da ação capilar em 1802 e congelou a água usando uma bomba de ar em 1810, a primeira produção artificial de gelo. (c) Jean-Baptiste Biot (1774-1862): em 20 de agosto de 1804 colaborou com Gay-Lussac em uma viagem de balão alcançando uma altura de 7.400 metros um recorde de ascensão para a época.[53]

A difusão de calor na matéria também foi estudada pelo físico e matemático francês Barão Jean Baptiste Fourier, Fig. 26(a), no ano de 1807. Na ocasião Fourier estudou a difusão do calor somente em corpos com formas bem especiais, tendo se dedicado no ano de 1811 à difusão do calor em corpos infinitos. No ano de 1822 Fourier mostrou a sua famosa equação da difusão do calor publicada no livro intitulado "Theorie analytique de la chaleur" (Teoria Analítica do Calor), obra que é considerada um marco na física-matemática.[54]

Devemos mencionar que entre os anos de 1807 e 1822 foram realizadas três importantes observações: *i*) no ano de 1809 Prévost verificou que quando dois corpos absorviam diferentes quantidades de calor, a sua emissão também era diferente, *ii*) a medição do fluxo de calor em barras metálicas realizadas em 1816 por Biot, e *iii*) a presença de linhas escuras no espectro solar observadas pelo físico alemão Joseph von Fraunhofer em 1814, as quais ficaram conhecidas como raias ou linhas de Fraunhofer, Fig. 26 (c). Essas linhas escuras podem ser observadas não só no espectro visível, mas também em toda faixa de radiação, de ondas de rádio a gama.[55]

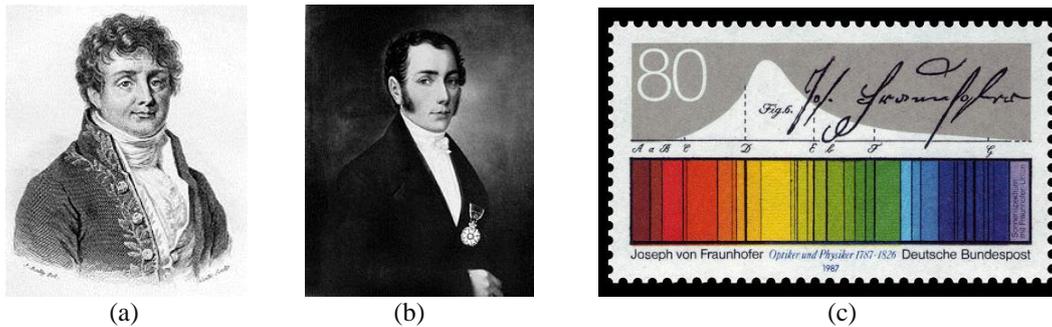

(a) (b) (c)

**Figura 26:** (a) Jean Baptiste Fourier (1768-1830): suas obras contribuíram muito para os fundamentos da termodinâmica e foi muito importante para a modelização matemática dos fenômenos físicos. (b) Joseph von Fraunhofer (1787-1826): produziu excelentes vidros ópticos e lentes objetivas acromáticas para telescópios. (c) Selo alemão comemora o 200º aniversário do nascimento de Fraunhofer (Alemanha, 1987).[56]

Apesar dos estudos acima mencionados, a ideia do calórico continuava sendo utilizada por vários cientistas para explicar diversos fenômenos térmicos. Por exemplo, entre os anos de 1817 e 1820, o reverendo J. B. Emmett propôs que o átomo era envolvido por uma atmosfera de calórico e em 1824 o engenheiro, físico e matemático francês Nicolas Léonard Sadi Carnot, Fig. 27(a), postulou que em máquinas térmicas o calórico era transformado em "força mecânica". Carnot publicou seus resultados em sua única obra: "Réflexions sur la Puissance Motrice du Feu et sur les Machines Propres a Développer Cette Puissance" (Reflexões sobre Potência Motriz do Fogo e Máquinas Próprias para Aumentar essa Potência).[57]

Importantes experiências realizadas entre 1842 e 1843 pelo médico inglês Julius Robert Mayer, Fig. 27(b), e pelo físico, também inglês, James Prescott Joule, Fig. 27(c), determinaram o que seria mais tarde conhecido como o "equivalente mecânico do calor". Cabe aqui relembrar, como mencionado na seção anterior, que Rumford em seus experimentos realizados entre 1787 e 1798 já havia feito uma associação entre calor e energia mecânica, precursora do equivalente mecânico do calor. Deve-se ressaltar também sobre a existência de notas escritas por Carnot no ano de sua morte, em 1832, em que procurava comparar a "queda" do calórico em sua máquina térmica com a queda da água de uma caixa com água.[58]

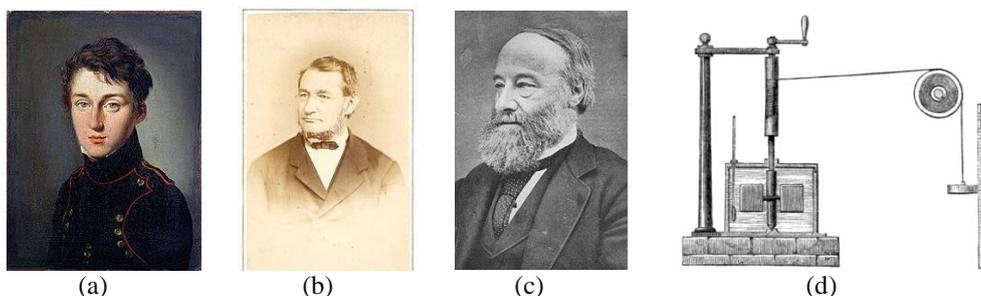

(a) (b) (c) (d)

**Figura 27:** (a) Nicolas Sadi Carnot (1796-1832): iniciou sua investigação sobre as propriedades dos gases após interessar-se pelos problemas industriais. (b) Julius Robert Mayer (1814-1878): um dos fundadores da termodinâmica junto com Kelvin e Joule. (c) James Prescott Joule (1818-1889): estudou a natureza do calor e descobriu relações com o trabalho mecânico. A nomenclatura joule, em sua homenagem para unidade de trabalho no SI só veio após sua morte. Joule trabalhou com Lorde Kelvin, para desenvolver a escala absoluta de temperatura. (d) Aparato de Joule para a medição do equivalente mecânico do calor, em que o trabalho realizado pelo peso de um objeto em queda é convertido em calor transferido à água.[59]

Em 1843, analisando o atrito entre diferentes materiais, o engenheiro civil e físico dinamarquês Ludwig August Colding, Fig. 28(a), concluiu que: "em todos os fenômenos naturais só se troca a forma de energia". Este enunciado pode ser considerado um conceito inicial da "Lei da Conservação da Energia". Em seus experimentos Colding conseguiu determinar o "equivalente mecânico do calor J". Mais tarde, em 1847, o físico e fisiologista alemão Hermann von Helmholtz, Fig. 28(b), baseando-se nas experiências precedentes de Mayer, Joule e Colding completou essa formulação inicial da lei de conservação da energia sendo reconhecida hoje como a "Primeira Lei da Termodinâmica". Em seguida, em 1865, destacam-se os trabalhos do físico e matemático alemão Rudolf Julius Emanuel Clausius, Fig. 28(c), que trouxe uma maior compreensão a respeito do processo de produção de trabalho na máquina de Carnot. Em 1865 Clausius introduziu o conceito de entropia, a qual recebeu uma intepretação probabilística no ano de 1877 pelo físico austríaco Ludwig Edward Boltzmann[60], o qual passou um longo tempo construindo e reconstruindo essa ideia probabilística da entropia. Essa interpretação dada por Boltzmann não imediatamente aceita levando um bom tempo até ser aceita pela comunidade científica da época. Ainda em 1865 Clausius apresentou uma formulação para as leis da termodinâmica, a saber:

- Primeira lei: *A energia do Universo é constante.*
- Segunda lei: *A entropia do Universo tende para um máximo.*

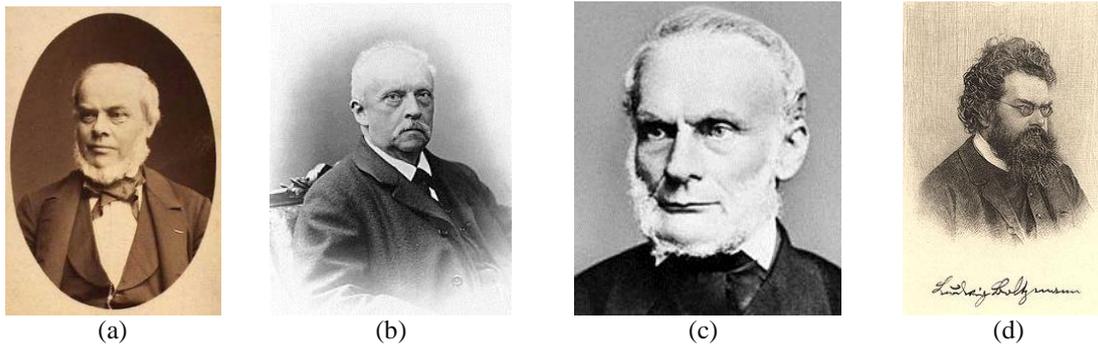

**Figura 28:** (a) Ludwig Colding (1815-1888): articulou o princípio da conservação de energia simultaneamente e independentemente com Joule e Mayer. (b) Hermann Helmholtz (1821-1894): criador da teoria da Panspermia Cósmica. (c) Rudolf Clausius (1822-1888): considerado um dos fundadores centrais da ciência da termodinâmica. (d) Ludwig Boltzmann (1844-1906): é considerado junto com Josiah Willard Gibbs e James Clerk Maxwell como o fundador da mecânica estatística; foi defensor da teoria atômica, numa época em que esta ainda era bem controversa.[61]

# RADIAÇÃO TÉRMICA

O físico e químico dinamarquês Hans Christian Oersted, Fig. 29(a), no ano de 1823, foi um dos primeiros interessados em medir a radiação térmica. No entanto, em sua época ainda não existiam aparelhos capazes de realizar tais medidas com precisão[62]. Avanços na medição da radiação térmica foram obtidos com a invenção do bolômetro, em um estágio ainda bastante primitivo, no ano de 1851, pelo físico sueco Adolf Ferdinand Svanberg (1806-1857). O bolômetro é uma espécie de termômetro para realizar medidas de radiação térmica. Em 1858 o físico escocês Balfour Stewart, Fig. 29(b), executando experiências sobre emissão e absorção de radiação térmica em uma placa de sal de rocha, afirmou que: o poder de emissão de cada tipo de substância é igual ao seu poder de absorção, para cada comprimento de onda do calor radiante[63]. Dentre as mais destacadas publicações de Stewart estão: "Observações com um espectroscópio rígido, aquecimento de um disco por movimento rápido no vácuo", "Equilíbrio térmico em um invólucro contendo matéria em movimento visível" e "Radiação interna em cristais uniaxiais".

Em 1859 o físico alemão Gustav Robert Kirchhoff, Fig. 29(c), realizou experiências sobre a absorção "$a$" e a emissão "$e$" do calor de materiais. Nestas experiências Kirchhoff deduziu que a relação entre a absorção e a emissão é uma função do comprimento de onda $\lambda$ da radiação absorvida ou emitida por um corpo que se encontra em uma temperatura $T$. Kirchhoff foi também o responsável pela introdução do conceito de corpo negro no ano de 1860. Corpo negro é todo corpo que absorve toda a radiação incidente nele. Assim para um corpo negro tem-se que: $e/a = 1$.[64]

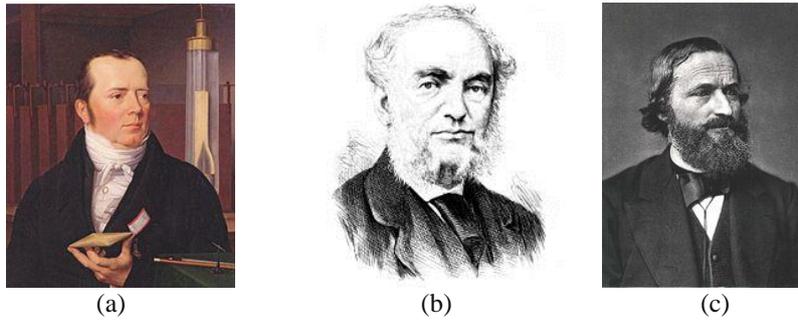

(a) (b) (c)

**Figura 29:** (a) Hans Oersted (1777-1851): descobriu que correntes elétricas podem criar campos magnéticos. (b) Balfour Stewart: fez pesquisas em meteorologia e campo magnético terrestre. (c) Gustav Kirchhoff (1824-1887): suas principais contribuições científicas foram no campo dos circuitos elétricos, na espectroscopia, na emissão de radiação dos corpos negros e na teoria da elasticidade.[65]

Em 1860, o matemático e físico britânico James Clerk Maxwell, Fig. 30(a), demonstrou que as velocidades das moléculas de um gás são distribuídas segundo a lei das distribuições dos erros, que foi formulada no ano de 1795 pelo matemático, físico e astrônomo alemão John Karl Friedrich Gauss, Fig. 30(b). Nesta lei a energia cinética das moléculas é proporcional à temperatura absoluta $T$ do gás. O físico austríaco Ludwig Eduard Boltzmann, Fig. 30(c), generalizou esta lei no ano de 1872, sendo atualmente conhecida como "lei de Maxwell-Boltzmann".[66]

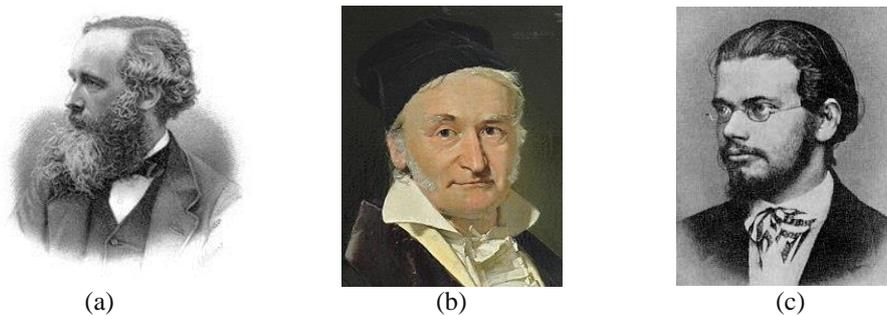

(a) (b) (c)

**Figura 30:** (a) James Maxwell (1831-1879): mais conhecido por ter dado forma final à teoria moderna do eletromagnetismo, que une a eletricidade, o magnetismo e a óptica. (b) Friedrich Gauss (1777-1855): é referido como "o príncipe da matemática". (c) Ludwig Boltzmann (1844-1906) aos 24 anos de idade: Foi defensor da teoria atômica numa época em que esta ainda era bem controversa.[67]

Maxwell demonstrou, em 1865, que um distúrbio eletromagnético num meio uniforme se propagava como uma onda. Posteriormente estas ondas foram chamadas de "ondas eletromagnéticas". Maxwell demonstrou também que a velocidade de propagação dessas ondas eletromagnética seria igual a velocidade da luz no vácuo. Estes resultados e a teoria eletromagnética desenvolvida por Maxwell foram publicados em 1873 em seu livro intitulado "Tratado sobre Eletricidade e Magnetismo". A confirmação experimental da teoria de Maxwell só foi concretizada mais tarde com a produção das primeiras ondas eletromagnéticas, no ano de 1887, pelo físico alemão Heinrich Rudolf Hertz, Fig. 31(a). Em homenagem a Hertz estas ondas eletromagnéticas foram chamadas de "ondas hertzianas".[68]

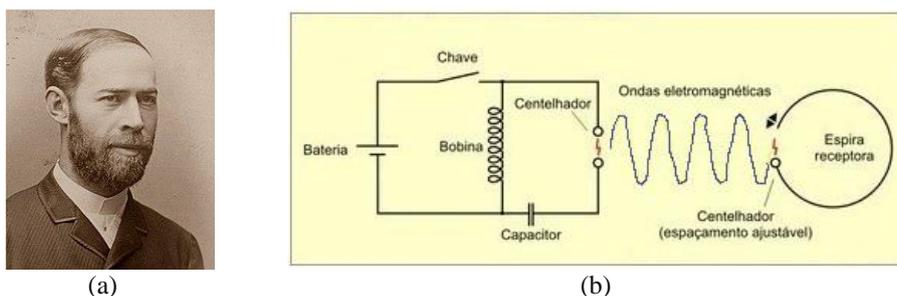

(a) (b)

**Figura 31:** (a) Heinrich Hertz (1857-1894): demonstrou experimentalmente a existência de radiação eletromagnética tal como previsto teoricamente por Maxwell. (b) Esquema do experimento de Hertz.[69]

Em 1879 o físico austríaco Josef Stefan, Fig. 32(a), obteve um importante resultado em relação à radiação térmica. Stefan fez medições da área sob as curvas do espectro radiante dos resultados que obteve realizando experiências para determinar a velocidade com que os corpos se esfriam[70]. Empiricamente Stefan obteve, por meio dessas medidas, a seguinte lei:

$$R \propto T^4,$$

onde $T$ é a temperatura do corpo e $R$ é a energia por unidade de tempo por unidade de área, ou seja, a potência pela área a qual é chamada de intensidade da radiação emitida por um corpo em uma dada temperatura $T$. Em uma notação matemática atual, a lei de Stefan pode ser escrita como:

$$R = \int_0^\infty I(\lambda, T)\, d\lambda,$$

onde $I(\lambda, T)$ era na época uma função desconhecida a qual dependia do comprimento de onda $\lambda$ da radiação e da temperatura $T$ do corpo. Então, o desafio dos cientistas na época era o de determinar qual era esta função $I(\lambda, T)$. Como já escrevemos acima esta função já havia sido mencionada em 1859 por Kirchhoff.

No ano de 1878, um ano antes do anúncio da lei de Stefan, o físico alemão Eugen Cornelius Joseph von Lommel, Fig. 32(b), foi um dos primeiros a tentar determinar qual era esta função $I(\lambda, T)$. Lommel não obteve sucesso uma vez que ele utilizou um modelo baseado em mecânica clássica para descrever as vibrações térmicas de um sólido[71]. Mesmo não tendo conseguido determinar qual era a função $I(\lambda, T)$ é importante citar Lommel, pois, sendo um dos primeiros a tentar determiná-la utilizando mecânica clássica começou a ser notado que a mecânica clássica não estava tendo êxito na explicação do fenômeno, algo a mais seria preciso.

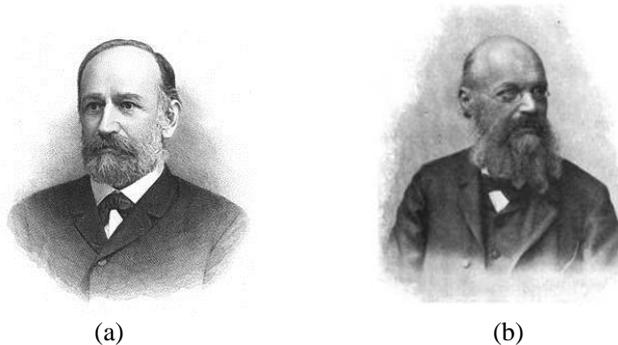

(a)          (b)

**Figura 32:** (a) Josef Stefan (1835-1893): ensinou física na Universidade de Viena, foi vice-presidente da Academia de Ciências de Viena e membro de várias instituições científicas da Europa. (b) Eugen Lommel (1837-1899): conhecido pelo polinômio de Lommel e a função de Lommel; foi orientador de Johannes Stark, Nobel de Física de 1919.[72]

No ano de 1884, Boltzmann foi um dos primeiros a demonstrar matematicamente a lei de Stefan. Para isto Boltzmann considerou que a radiação eletromagnética no interior de um corpo negro era análoga a um gás[73]. Em sua demonstração Boltzmann utilizou a teoria eletromagnética de Maxwell associada à termodinâmica de Carnot. Neste caminho Boltzmann demostrou que

$$R = \sigma T^4,$$

onde $\sigma$ é uma constante que posteriormente passou a ser chamada de constante de Stefan-Boltzmann. Nota-se, no entanto, que a demonstração de Boltzmann foi realizada sem se conhecer qual era a função $I(\lambda, T)$.

Dois anos depois, em 1886, o físico e astrônomo norte-americano Samuel Pierpont Langley, Fig. 33(a), começou a realizar medidas do espectro da radiação do corpo negro com o bolômetro que ele próprio havia inventado no ano de 1878. Langley mediu a intensidade da radiação eletromagnética solar na faixa de vários comprimentos de onda, incluindo a faixa do espectro no infravermelho. Langley verificou um deslocamento do máximo da intensidade de cada radiação com o aumento de temperatura. Langley realizou também medições da radiação térmica emitida por um pedaço de cobre em altas temperaturas[74]. No ano de 1893, esta observação experimental feita por Langley foi matematicamente demonstrada pelo físico alemão Wilhelm Wien, Fig. 33(b). Este resultado posteriormente passou a ser chamado de lei do deslocamento de Wien[75], que pode ser expressa da seguinte forma:

$$T \cdot \lambda_{max} = \text{constante}.$$

Esta lei trouxe ainda mais dificuldades para os cientistas da época que tentavam obter a função $I(\lambda, T)$.

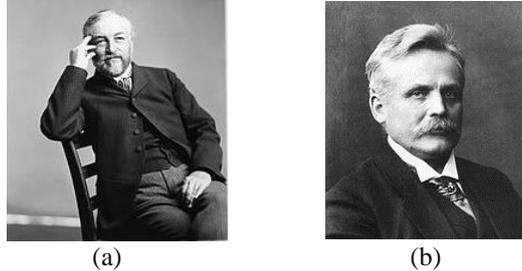

(a)            (b)

**Figura 33:** (a) Samuel Pierpont Langley (1834-1906): publicou em 1890 suas observações sobre os infravermelhos junto com Frank Washington Very, sendo posteriormente a base para a formulação dos primeiros cálculos sobre o efeito estufa, realizados por Svante Arrhenius. (b) Wilhelm Wien (1864-1938): trabalhou no laboratório de Hermann von Helmholtz e, em 1886, recebeu seu Ph.D. com uma tese sobre a difração da luz em cima de metais e sobre a influência de diversos materiais sobre a cor da luz refratada.[76]

No ano de 1896 foi realizada a primeira determinação empírica da função $I(\lambda, T)$. Esta determinação foi feita de forma independente por Wilhelm Wien e pelo físico, também alemão, Louis Karl Heinrich Friedrich Paschen, Fig. 34(a). Ambos chegaram a seguinte expressão:

$$I(\lambda, T) = \frac{c_1}{\lambda^5 e^{c_2/\lambda T}},$$

onde $c_1$ e $c_2$ são constantes[77]. Ressaltamos também aqui o trabalho do físico russo Vladimir Alexandrovich Michelson, Fig. 34(b), que em 1888, usou a distribuição de velocidades de Maxwell para tentar determinar a função $I(\lambda, T)$. Michelson foi o primeiro a aplicar métodos de física estatística para determinar a função da distribuição de energia no espectro de emissão do corpo negro.

Um importante trabalho teórico para o entendimento do espectro de radiação térmica foi realizado, no ano de 1897, pelo matemático e físico inglês Sir Joseph Larmor, Fig. 34(c). Larmor demostrou que uma carga elétrica acelerada irradia ondas eletromagnéticas. Dessa forma, a radiação térmica emitida por um corpo teria como origem a existência de cargas elétricas aceleradas pelo processo de agitação térmica na superfície de um corpo, e o espectro de radiação contínuo emitido por um corpo em uma dada temperatura seria devido ao modo aleatório como as cargas superficiais estão sendo aceleradas. Larmor foi o primeiro a publicar a transformação de Lorentz, em 1897, dois anos antes de Hendrik Antoon Lorentz e oito anos antes de Albert Einstein[78].

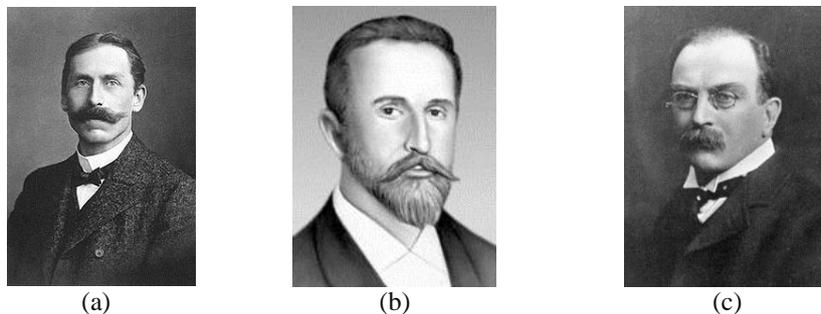

(a)            (b)            (c)

**Figura 34:** (a) Friedrich Paschen (1865-1940): conhecido por seus trabalhos sobre descargas elétricas. (b) Vladimir Alexandrovich Michelson (1860-1927): professor titular da cadeira de física e meteorologia no Instituto Agrícola de Moscou. (c) Sir Joseph Larmor (1857-1942): em 1903, foi nomeado Professor Lucasiano de Matemática em Cambridge, cargo que manteve até sua aposentadoria em 1932.[79]

As primeiras medidas mais precisas da intensidade do espectro de radiação de corpo negro em função do comprimento de onda foram realizadas em novembro de 1899, Fig. 35, pelo físico alemão Otto Richard Lummer, Fig. 36(a), e pelo físico e médico também alemão Ernst Pringsheim, Fig. 36(b). Eles utilizaram um instrumento que era basicamente igual aos espectrômetros de prisma usados nas medidas dos espectros ópticos, diferindo apenas nos materiais especiais que eram necessários para que as lentes e prismas fossem transparentes à radiação térmica de frequência relativamente baixa[80]. Nota-se claramente nas curvas apresentadas na Fig. 35 um deslocamento para a esquerda do pico (máximo) de cada curva para cada uma das temperaturas indicadas na figura. Este é o chamado "deslocamento de Wien"[81].

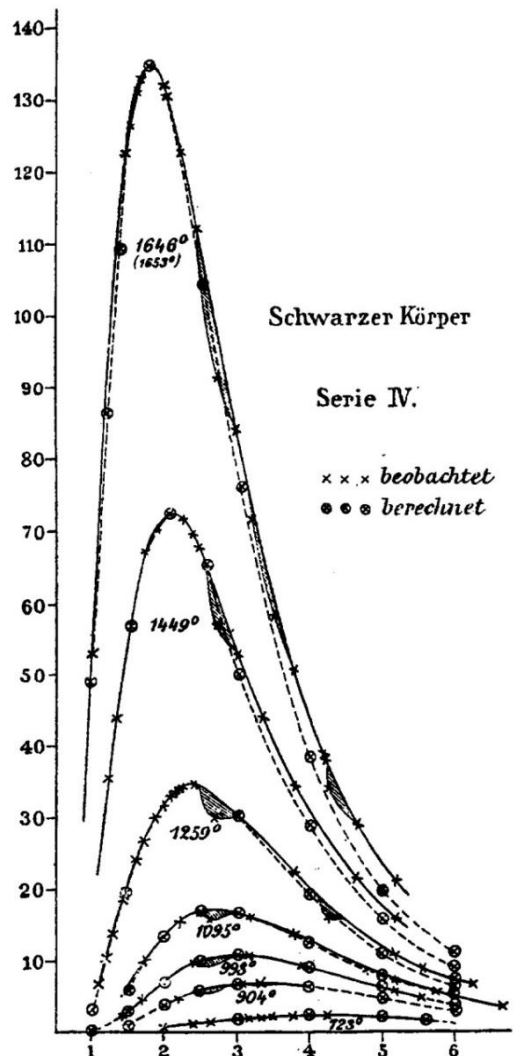

**Figura 35**: Intensidade espectral como função do comprimento de onda obtida por Lummer e Pringsheim em novembro de 1899.[82]

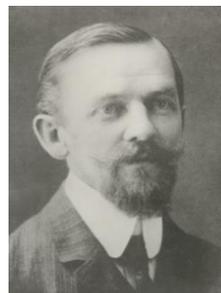 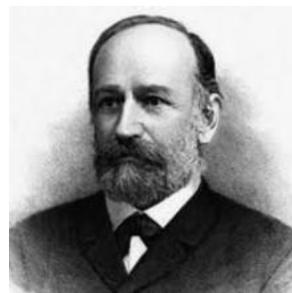

(a)          (b)

**Figura 36:** (a) Otto Richard Lummer (1860-1925): físico alemão; realizou estudos sobre luminotecnia e o espectro do corpo negro; estudou em diversas universidades da Alemanha, sendo em 1884 assistente de Hermann von Helmholtz em Berlim; lá trabalhou na Physikalisch-Technische Bundesanstalt, onde foi professor a partir de 1894. Em 1904 tornou-se professor em Breslau. (b) Ernst Pringsheim (1859-1917): médico e físico alemão; se especializou em pesquisas de física teórica e experimental relativas ao calor e à radiação luminosa; é conhecido por ser o primeiro a desenvolver o radiômetro, um instrumento útil para medir a radiação infravermelha.[83]

O físico e matemático inglês John William Strutt, também conhecido como Lord Rayleigh, Fig. 37(a), continuou a busca pela função $I(\lambda, T)$ que se ajustasse aos resultados experimentais[84]. Rayleigh observou, em junho de 1900, que a fórmula de Paschen-Wien só era válida para pequenos comprimentos de onda. Rayleigh chegou à seguinte expressão:

$$I(\lambda, T) = \frac{c_3 T}{\lambda^4 e^{c_2/\lambda T}},$$

onde $c_2$ e $c_3$ são constantes. No entanto, dois físicos alemães, Heinrich Leopold Rubens, Fig. 37(b), e Ferdinand Kurlbaum, Fig. 37(c), mostraram de forma experimental, em outubro de 1900, que essa fórmula de Rayleigh só valia para grandes comprimentos de onda. Dessa maneira, a fórmula de Rayleigh divergia para pequenos comprimentos de onda, ou seja, para altos valores de frequências da radiação emitida[85]. Esta divergência entre a fórmula teórica de Rayleigh com os resultados experimentais para altos valores de frequência foi chamada de "catástrofe do ultravioleta".

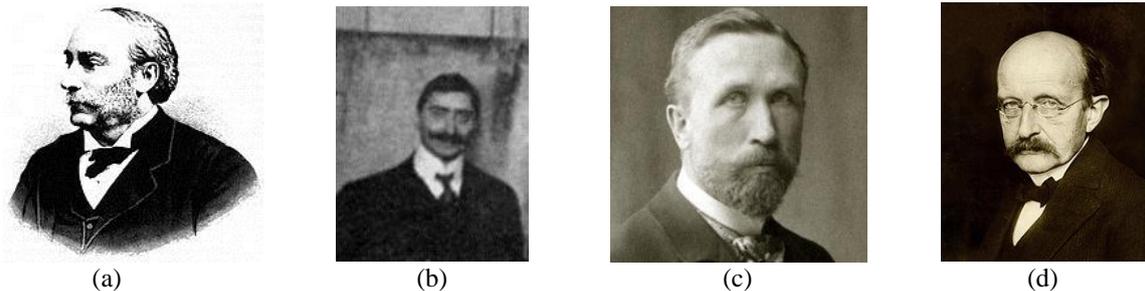

(a)  (b)  (c)  (d)

**Figura 37:** (a) Lord Rayleigh (1842-1919): recebeu o Nobel de Física, em 1904, por pesquisas sobre a densidade de gases e pela descoberta do argônio, em pesquisa realizada em conjunto com o químico inglês Sir William Ramsay. (b) Heinrich Rubens (1865-1922): obteve seu doutorado em agosto de 1889 na Universidade Humboldt, sob a orientação de Kundt, com a tese *Die seletivo reflexion der metalle* (O reflexo seletivo de metais). (c) Ferdinand Kurlbaum (1857-1927): estudou matemática e física em Heidelberg e em Berlim com Hermann Helmholtz; em 1887 concluiu sua dissertação sobre a determinação do comprimento de onda das linhas de Fraunhofer. (d) Max Karl Planck (1858-1947): considerado o pai da física quântica e um dos físicos mais importantes do século XX.[86]

No dia 19 de outubro de 1900 o físico alemão Max Karl Ernst Ludwig Planck, Fig. 37(d), mostrou uma solução para esse impasse. Neste dia, Planck apresentou um trabalho à Sociedade de Física de Berlim em que analisava as fórmulas para a curva de radiação do corpo negro propostas por Rayleigh e por Paschen-Wien[87]. Como a fórmula de Rayleigh se ajustava à curva experimental para a radiação de corpo negro para grandes comprimentos de onda e a fórmula de Paschen-Wien se ajustava à curva experimental para pequenos comprimentos de onda, Max Planck fez uma interpolação entre as duas obtendo a seguinte fórmula:

$$I(\lambda, T) = \frac{c_1}{\lambda^5 (e^{-c_2/\lambda T} - 1)},$$

onde $c_1$ e $c_2$ são constantes. Essa fórmula obtida por Planck se reduz à fórmula de Rayleigh quando $\lambda$ é grande e à fórmula de Paschen-Wien quando $\lambda$ é pequeno.

Sabe-se que inicialmente em sua demonstração Planck usou todos os recursos disponíveis da Termodinâmica pré-Boltzmann, mas sem sucesso é claro. E seguindo por esse caminho clássico Planck não encontrou nenhum erro nos cálculos realizados por Lord Rayleigh. Planck adotou, então, a interpretação probabilística da entropia proposta por Boltzmann em 1877. No entanto, para obter o resultado desejado, Planck supôs em sua demonstração que a energia dos osciladores moleculares das paredes do corpo negro, que vibravam com uma frequência $\nu$, variava de forma discreta, da seguinte forma: $\mathcal{E} = h\nu$, onde $h$ é uma constante.

Planck achava que essa suposição era somente um artifício matemático, e que a constante $h$ desaparecia durante os cálculos ao se tomar $h \to 0$. No entanto, Planck constatou que só era possível obter uma fórmula que se ajustava aos resultados experimentais se a constante $h$ tivesse um valor finito, ou seja, não podia tender a zero. Seguindo esse caminho, Planck apresentou à Sociedade de Física de Berlim, no dia 14 de dezembro de 1900, a sua demonstração na qual apresentou os valores das constantes $c_1$, $c_2$ e $h$: $c_1 = hc^2$, $c_2 = hc/k$, $h = 6{,}63 \times 10^{-34}$ J.s, onde $k$ é a constante de Boltzmann e $c$ é a velocidade da luz no vácuo. O valor de $h$ determinado por Planck foi o melhor valor por ele obtido para que a sua teoria se ajustasse o melhor possível aos dados experimentais.

Deve-se ressaltar que, apesar do resultado satisfatório obtido por Planck no que diz respeito ao bom ajuste de sua equação com os resultados experimentais, a hipótese adotada por Planck de uma "variação discreta da energia" na dedução de sua fórmula não foi inicialmente muito bem aceita pela comunidade científica. Até mesmo o próprio Planck tinha uma certa ressalva quanto à sua hipótese. Assim os físicos da época continuaram utilizando a Física Clássica para tentar obter a função $I(\lambda, T)$ sem utilizar a ideia revolucionária de discretização da energia adotada por Planck, ou seja, continuariam utilizando nesta tentativa a variação contínua da energia usada na Física Clássica. Assim, utilizando a Física Clássica Lord Rayleigh obteve, em maio de 1905, a seguinte expressão: $I(\lambda, T) = 64\pi kT/\lambda^4$. Porém, havia um pequeno erro cometido na dedução dessa expressão. O físico britânico Sir James Hopwood Jeans, Fig. 38(a), corrigiu o erro na fórmula de Rayleigh e pouco tempo depois, em julho de 1905, apresentou a célebre fórmula: $I(\lambda, T) = 8\pi kT/\lambda^4$, a qual ficou conhecida como "fórmula de Rayleigh-Jeans".[88]

Deve-se aqui salientar que ainda no ano de 1905, o físico Albert Einstein, Fig. 38(b), utilizou a hipótese de discretização da energia feita por Planck para explicar o efeito fotoelétrico[89]. Assim, o efeito fotoelétrico se tornou uma forma experimental independente da radiação de corpo negro para determinação da constante $h$, constatando-se que a constante $h$ não era simplesmente um artifício matemático para obtenção da fórmula de Planck. O responsável pela determinação experimental da constante $h$ foi o físico americano Robert Andrews Millikan, Fig. 38(c), que em 1914 encontrou o valor $h = 6{,}57 \times 10^{-34}$ J. s, valor este muito próximo do encontrado por Planck em 1900. Citando as palavras de Millikan[90]: "*O efeito fotoelétrico fornece uma prova independente da apresentada pela radiação de corpo negro, da exatidão da hipótese fundamental da teoria quântica, ou seja, a hipótese da emissão contínua ou explosiva da energia que é absorvida das ondas pelos constituintes eletrônicos dos átomos. Ele materializa, por assim dizer, a quantidade $h$ descoberta por Planck em seu estudo da radiação de corpo negro e, como nenhum outro fenômeno, nos faz acreditar que o conceito físico básico que está por trás do trabalho de Planck corresponde à realidade.*"

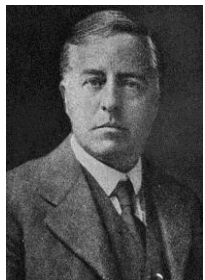 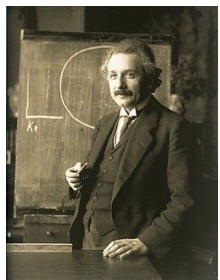 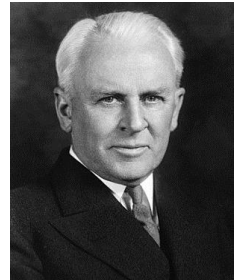

(a)          (b)          (c)

**Figura 38:** (a) Sir James Jeans (1877-1946): uma de suas descobertas mais importantes é o comprimento de Jeans, que é o raio crítico de uma nuvem interestelar no espaço o qual depende da massa, tamanho e densidade da nuvem; uma nuvem menor que o comprimento de Jeans não terá gravidade suficiente para superar as forças de gases exógenos, enquanto que uma nuvem maior que o dito comprimento se colapsará numa estrela. (b) Albert Einstein (1879-1955): foi laureado com o Prêmio Nobel de Física de 1921 por suas contribuições à física teórica e por sua descoberta da lei do efeito fotoelétrico, fundamental no estabelecimento da teoria quântica. (c) Robert Andrews Millikan (1868-1953) em 1920: foi o primeiro cientista a determinar o valor da carga do elétron através da "experiência de Milikan".[91]

A "realidade" da constante $h$ continuou sendo verificada em experimentos posteriores, como o efeito Compton em 1923 e em experimentos de produção de fótons como a radiação X.[92]

# CONSIDERAÇÕES FINAIS

Foi possível com base nesta pesquisa, discutir e ordenar cronologicamente as ideias sobre o conceito fundamental de calor e radiação térmica, transcorrendo sobre conceitos primordiais até a física moderna. Foi realizada uma apresentação dos principais cientistas e os seus feitos quanto ao tema em questão. Um conhecimento a respeito da evolução histórica de conceitos físicos que foram se transformando ao longo do tempo é condição fundamental para um bom entendimento sobre o assunto em questão, uma vez que a ligação cronológica dos fatos coloca degraus, ou seja, preenche lacunas existentes na formação daquele conceito.

Ressalta-se que a utilização de história no ensino de ciências deve ser feita com muita cautela. É importante discutir o contexto da época e principalmente a falta de documentação adequada. Devido à falta de documentos da época que pudessem esclarecer o que realmente ocorreu, muitas vezes têm-se vários questionamentos[93] como, por exemplo, qual foi a real metodologia utilizada por um cientista na explicação de um determinado fenômeno ou experimento[94]. Mesmo com estas questões, acredita-se no valor dos fatos históricos na facilitação da aprendizagem em diversos tópicos de física e de ciências em geral[95].

# REFERÊNCIAS

[27] H. Cavendish, *The Scientific Papers of the Honourable Henry Cavendish* (New York: Cambridge University Press, 2011).
[28] Fonte: <https://pt.wikipedia.org/wiki/Henry_Cavendish> (acessado em 10 de maio de 2021).
[29] Madison Smartt Bell, *Lavoisier in the Year One* (New York: Atlas Books, 2005).
[30] Fonte: <https://pt.wikipedia.org/wiki/Claude_Berthollet> (acessado em 28 de maio de 2021).
[31] Fonte: <https://br.pinterest.com/SciHistoryOrg/mysteries-of-chemistrys-most-famous-people/> (acessado em 28 de maio de 2021).
[32] Guilherme W. Tavares, Alexandre G. S. Prado, *Calorímetro de gelo: uma abordagem histórica e experimental para o ensino de química na graduação*, Química Nova, vol. 33, n. 9, 1987-1990 (2010).
[33] Fonte: <https://www.chemistryworld.com/opinion/laplaces-calorimeter/1010082.article> (acessado em 29 de maio de 2021).
[34] S. S. Kutateladze, *The Mathematical Background of Lomonosov's Contribution*, <http://www.math.nsc.ru/LBRT/g2/english/ssk/mvl_e.html> (acessado em 29 de maio de 2021).
[35] Giordano Repossi, *Ciência e Revolução. A Química. O Mundo Misterioso da Molécula* (Lisboa: Círculo de Leitores, 1977).
[36] John Gribbin, *Science, a History* (London: Penguin Books, 2003).
[37] Fonte: <https://pt.wikipedia.org/wiki/Amedeo_Avogadro> (acessado em 28 de maio de 2021).
[38] John J. O'Connor and Edmund F. Robertson, <https://mathshistory.st-andrews.ac.uk/Biographies/Lagrange/> (acessado em 28 de maio de 2021).
[39] John L. Heilbron, René Sigrist, *Jean-André Deluc, historian of earth and man* (Geneva: Slatkine, 2011).
[40] <https://ao.regionkosice.com/wiki/Joseph_Black#cite_note-9> (acessado em 28 de maio de 2021).
[41] Arthur Donovan, *James Hutton, Joseph Black and the Chemical Theory of Heat*, Ambix, vol. 25, n. 3, 176-190 (1978).
[42] Miguel Spinelli, *Bacon, Galileu e Descartes. O renascimento da filosofia grega* (São Paulo: Loyola, 2013).
[43] Fonte: <https://pt.wikipedia.org/wiki/Francis_Bacon> (acessado em 28 de maio de 2021).
[44] Benjamin Thompson, *New Experiments upon Gun-Powder, with Occasional Observations and Practical Inferences*, Philosophical Transactions of the Royal Society of London, vol. 71, 229-328 (1781).
[45] Fonte: <https://pt.wikipedia.org/wiki/Benjamin_Thompson> (acessado em 28 de maio de 2021).
[46] Fonte: <https://docplayer.com.br/76256408-Calor-trabalho-e-1-a-lei-da-termodinamica.html> (acessado em 28 de maio de 2021).
[47] Humphry Davy, *Some Observations and Experiments on the Papyri Found in the Ruins of Herculaneum*, Philosophical Transactions, vol. 111, 191-208 (January, 1821).
[48] Pierre Prevost, *Du calorique rayonnant* (Paris: J.J. Paschoud, 1809).
[49] Fonte: <https://de.wikipedia.org/wiki/Pierre_Pr%C3%A9vost> (acessado em 28 de maio de 2021).
[50] Richard Holmes, *The age of wonder* (New York: Vintage Books, 2008).
[51] Fonte: <https://de.wikipedia.org/wiki/Wilhelm_Herschel> (acessado em 02 de junho de 2021).
[52] M. P. Crosland, *Biot, Jean-Baptiste. Dictionary of Scientific Biography*, v. 2 (New York: Charles Scribner's Sons, 1980).
[53] Fonte <https://de.wikipedia.org/wiki/Jean-Baptiste_Biot> (acessado em 02 de junho de 2021).
[54] Joseph Fourier, *The Analytical Theory of Heat* (Cambridge: reissued by Cambridge University Press, 2009).
[55] Myles W. Jackson, *Spectrum of Belief: Joseph von Fraunhofer and the Craft of Precision Optics* (Cambridge: MIT Press., 2000).
[56] Fonte: <http://www.astronoo.com/pt/artigos/espectroscopia.html> (acessado em 02 de junho de 2021).
[57] J. Srinivasan, *Sadi Carnot and the Second Law of Thermodynamics*, RESONANCE, vol. 6, n. 11, November, 42-48 (2001).
[58] H. S. Allen, *James Prescott Joule and the Unit of Energy*, Nature, vol. 152, n. 3856, 354 (1943).
[59] Fonte: <https://pt.wikipedia.org/wiki/Nicolas_L%C3%A9onard_Sadi_Carnot> (acessado em 02 de junho de 2021).
[60] R. Clausius, *On a Modified Form of the Second Fundamental Theorem in the Mechanical Theory of Heat*, Phil. Mag., vol. 12, n. 77, 81-98 (August, 1856).
[61] Fonte: <https://pt.wikipedia.org/w/index.php?title=Especial:Pesquisar&search=Ludwig+Colding&ns0=1> (acessado em 02 de junho de 2021).
[62] D. C. Christensen, *Hans Christian Orsted* (Oxford: Oxford University Press, 2013).
[63] Balfour Stewart, *The New International Encyclopedia*, *vol. 18* (New York: Dodd, Mead and Company, 1905).
[64] Thomas Hockey, *Kirchhoff, Gustav Robert. The Biographical Encyclopedia of Astronomers* (Berlin, Springer Nature, 2009).
[65] Fonte: <https://pt.wikipedia.org/wiki/Gustav_Kirchhoff > (acessado em 18 de junho de 2021).
[66] Ronaldo Rodrigues Pelá, <http://www.ief.ita.br/~rrpela/downloads/fis14/FIS14-2012-aula23.pdf>. (acessado em 18 de junho de 2021).
[67] Fonte: <https://pt.wikipedia.org/wiki/Ludwig_Boltzmann> (acessado em 18 de junho de 2021).